

\def\dvp{\raisebox{-.45ex}{\rlap{$
=$}} \raisebox{-.45ex}{$\hskip .48ex { |}$}}
\def\dvm{\raisebox{-.45ex}{\rlap{$=$}} }

\def\DP{{\scriptsize{\dvp}}~~}
\def\DM{{\scriptsize{\dvm}}~~}

\newcommand{\beq}{\begin{equation}}
\newcommand{\eeq}{\end{equation}}
\newcommand{\bea}{\begin{eqnarray}}
\newcommand{\eea}{\end{eqnarray}}

\input dnu.tex

\border\headpic {\hbox to\hsize{January 1994 \hfill {UMDPP 94-074}}}
{\hbox to\hsize{\hfill {QMW 93-35}}}
\par
\setlength{\oddsidemargin}{0.3in}
\setlength{\evensidemargin}{-0.3in}
\begin{center}
\vglue .08in
{\large\bf Superspace Supervortices}
\\[.72in]
{\large S. James Gates, Jr.\footnote{Research supported by NSF grant
NSF-PHY-91-19746}${}^,$\footnote{e-mail: gates@umdhep.umd.edu}}
\\[.02in]
{\it Physics Department\\
University of Maryland at College Park\\
College Park, MD 20742-4111, USA}\\[.02in] and \\[.02in]
{\it Mathematical Sciences Research Institute\\
Berkeley, California 94720, USA}\footnote{Research at MSRI supported
in part by NSF grant \#DMS 9022140} \\[.04in]
and\\[.04in]
{\large Oleg A. Soloviev\footnote{Work supported by
S.E.R.C.}${}^,$\footnote{e-mail: soloviev@V1.PH.QMW.ac.uk}}\\[.02in]
{\em Physics Department, Queen Mary and Westfield College, \\
Mile End Road, London E1 4NS, United Kingdom}\\ [1.4in]

{\bf ABSTRACT}\\[.002in]
\end{center}
\begin{quotation}
{We present the theory describing supersymmetrical vortices in the curved
superspace of the (1,0) supergravity. The action is defined as a (1,0)
locally supersymmetric $SU(2)/U(1)$ coset perturbed by the cosmological
constant-like term. The perturbation is such that it preserves the
integrability
of the coset model. Because of supersymmetry the perturbed theory is an
exact quantum system provided a proper dilaton is taken into account. The
exact value of the dilaton is determined in the supersymmetric case by the
quasi-classical background of the bosonic coset. }

\endtitle

\section{Introduction}

{}~~~It has long been known that vortices and strings share many surprising
similar features [1, 2]. Therefore, a number of attempts have been made
in order to clarify some properties of string models based on the theory of
vortices. For instance, a few years ago, several authors [3, 4] discussed the
implications of the planar vortex Kosterlitz-Thouless transition for string
theories compactified on a circle. Quite recently, it has been realized [5]
that
we can learn something interesting about vortices themselves by applying ideas
``borrowed" from string theory. In this paper, we are going to examine the
possible role that world sheet supersymmetry might play in explaining certain
quantum aspects of the vortex model discussed in [1, 5].

The important observation in [5] is that the classical vortex action derived in
[1] can be treated as the $SU(2)/U(1)$ coset perturbed by the first ``thermal"
operator. The perturbation was demonstrated to sustain the integrability of the
coset providing an infinite number of conserved currents related to
$W_\infty$-algebras [5]. The system was argued to describe a broad class of
integrable two dimensional models. Note in this connection
that the relation between the similar
coset ($SL(2)/U(1)$) and integrable models was also considered in [6].

In the Lagrangian formalism many coset models can be described as gauged
Wess-Zumino-Novikov-Witten (WZNW)
theories [7]. The last circumstance is the key to
ascribing stringy effects to vortex models. The aim of this paper is to
extend the interconnection between vortices and coset theories to
the case of supersymmetric coset models.

\section{Gauged (1,0) WZNW model}
{}~~~~It is well known that when we are
dealing with supersymmetrical models, it is more convenient conceptually and
technically to work in terms of superfields defined on a proper (curved)
superspace. Therefore, we start with a superspace description of the gauged
(1,0) locally supersymmetric WZNW model [8]. Note that (1,0) supersymmetry is
the simplest one of all known supersymmetries. We want the supersymmetry to be
local because we will show that (super)gravity plays a significant role at
the quantum level.  The superspace action of the (1,0) locally supersymmetric
gauged (level $k$) WZNW model is defined as follows [8]
\begin{equation}
S(g,\Gamma_+,\Gamma_{\DM})=S(g)\;+\;{ik\over\pi}\int d^3zE^{-1}Tr\{\nabla_\DM
gg^{-1}\Gamma_+\;-\;\Gamma_\DM
g^{-1}\nabla_+g\;+\;g^{-1}\Gamma_+g\Gamma_\DM\;-\;\Gamma_+\Gamma_\DM\},
\end{equation}
where $\Gamma_+,\;\Gamma_\DM$ are the gauge superfields valued in the adjoint
representation of ${\cal H}$ which is a subalgebra of the Lie algebra ${\cal
G}$ associated with the group Lie $G$ whose group elements $g$ are the matrix
(1,0) scalar superfields. A curved (1,0) superspace is parameterized by the
coordinates $z^M=(\sigma^m,\;\zeta^+)$. Here $\sigma^m=(\tau,\;\sigma)$ are
world sheet bosonic coordinates, $\zeta^+$ is the Grassman coordinate for the
(1,0) superspace. The Lorentz-covariant and super-covariant derivatives in
(1,0) superspace are:
\begin{equation}
\nabla_A=(\nabla_+,\;\nabla_\DP,\;\nabla_\DM)=E_A~^M\partial_M+\omega_A{\cal
M},
\end{equation}
where we have introduced the superzweibein $E_A~^M(z)$ and the supeconnection
$\omega_A(z)$ in (1,0) superspace. The quantity ${\cal M}$ is the generator of
Lorentz rotations. The $\nabla_A$ derivatives satisfy the algebra [9]
\begin{equation}
\{\nabla_+,\nabla_+\}=2i\nabla_\DP,\;\;\;\;\;\;\;[\nabla_\DM,\nabla_+]=2i
\Sigma^+{\cal M},\end{equation}
where we have introduced the superfield strength, $\Sigma^+$, of 2D (1,0)
supergravity. The (1,0) superspace density is $E^{-1}=[Ber(E_A~^M)]^{-1}$.
For more notation, refer to [8, 9].

The action given by eq. (1) is invariant under the transformations
\begin{eqnarray}
g&\to&\Lambda g\Lambda^{-1}, \;\;\;\Lambda\in H\subset G,\nonumber\\ & & \\
\Gamma_A&\to&\nabla_A\Lambda\Lambda^{-1}+\Lambda\Gamma\Lambda^{-1}.\nonumber
\end{eqnarray}

It is worth emphasizing that the gauge symmetry in (4) is not anomalous at the
quantum level despite the chiral structure of the (1,0) supersymmetric theory
[8]. To explain this, we define components of the relevant (1,0) superfields as
\begin{eqnarray}
\; \; \; g|&\equiv & u \; \;,\; \;(g^{-1}\nabla_+g)_{\cal M}|\equiv i\psi_+,\;
(g^{-1}\nabla_+g)_{\cal H}|\equiv i\beta_+ \; \; \; ,
\nonumber\\ & & \\
\nabla_+\Gamma_+|&\equiv& iV_\DP \; \; ,\; \; \Gamma_\DM|\equiv
V_\DM\; \; , \; \; \nabla_+\Gamma_\DM|\equiv\lambda_- \; \; \;. \nonumber
\end{eqnarray}
The component $\Gamma_+|$ can be set to zero by an algebraic supersymmetry
transformation that corresponds to a choice of the Wess-Zumino gauge. In
formulas (5), $|$ means taking the first component of a superfield or an
operator; $(...)_{{\cal H}({\cal M})}$ denotes a projection onto the subalgebra
${\cal H}({\cal M})$ of ${\cal G}$. The algebra ${\cal G}$ is supposed to have
the orthogonal decomposition [8]
\begin{equation}
{\cal G}={\cal H}\oplus{\cal M},\;\;\;[{\cal H},{\cal H}]\subset{\cal
H},\;\;\;[{\cal H}, {\cal M}]\subset{\cal M}.\end{equation}
Then, in terms of components, in the supeconformal gauge the action takes the
form
\begin{equation}
S(g,\Gamma_+,\Gamma_\DM)=S_{WZNW}(u,V_\DP,V_\DM)-{i\over\pi}\int d^2\sigma\{
Tr\psi_+\partial_\DM\psi_++Tr[\beta_+(\partial_\DM-2V_\DM)\beta_++i\lambda_-
\beta_+]\},\end{equation}
where the $S_{WZNW}$ denotes the gauged (level $k$) bosonic WZNW
action\footnote{Note that our definition of component fields in (5)
assumes that the level $k$ is that corresponding to the direct product of the
bosonic affine algebra and the free fermionic algebra. So that there is no
extra shift in the level due to quantum effects.}.

A potential chiral anomaly may seem to arise from the third term in eq. (7),
since this describes the gauged Majorana-Weyl 2D spinors. Fortunately, there is
the fourth term (the last one) in eq. (7) which clearly ensures a nondynamical
nature of $\beta_+$ and $\lambda_-$ in the adjoint representation of ${\cal
H}$, because these spinors vanish on shell. So, the anomalies due to the chiral
spinors in our theory never arise and we can freely gauge out $H$-valued
degrees of freedom both at the classical and the quantum levels.

{}From now on we will focus our attention on the $SU(2)/U(1)$ coset model. In
other words we choose $G=SU(2)$ and $H=U(1)$. Note that in general we can gauge
the $U(1)$ subgroup of $SU(2)$ in many ways. Henceforth we will think of $U(1)$
as the Cartan subgroup of $SU(2)$ so that the decomposition in (6) is
fulfilled. We fix the gauge by choosing $SU(2)$ group element of the form
\begin{equation}
g=\left(\begin{array}{cc}
g_0+ig_3 & ig_1\\
{}~&~\\
ig_1& g_0-ig_3\\
\end{array}\right),\end{equation}
at the condition
\begin{equation}
g_0^2\;+\;g^2_1\;+\;g^3_3=1,\end{equation}
where $g_0,\;g_1,\;g_3$ are real (1,0) superfields. In the adopted gauge, the
(1,0) superspace action (1) can be written in terms of two complex scalar
superfields
\begin{equation}
u=g_0+ig_3,\;\;\;\;\;\bar u=g_0-ig_3,\;\;\;\;u\bar u\le1\end{equation}
as follows
\begin{equation}
S={ik\over2\pi}\int d^3zE^{-1}{(\nabla_+u\nabla_\DM\bar u\;+\;\nabla_+\bar
u\nabla_\DM u)\over1-u\bar u}.\end{equation}
The action yields the following equations of motion in (1,0) superspace
\begin{eqnarray}
\nabla_+\nabla_\DM\bar u\;+\;{u\nabla_\DM\bar u\nabla_+\bar u\over1-u\bar
u}&=&0,\nonumber\\ & & \\
\nabla_+\nabla_\DM u\;+\;{\bar u\nabla_\DM u\nabla_+ u\over1-u\bar u}&=&0,
\nonumber\end{eqnarray}
which are the (1,0) supersymmetric generalization of the equations obtained for
the bosonic $SU(2)/U(1)$ ($SL(2)/U(1)$) coset [1, 5]([6]). Since the bosonic
model is completely integrable [1, 2] its supersymmetric generalization
enjoys this same property. In order to construct supercovariant conserved
currents it is useful to define supercovariant Lax operators. Let us define the
following pair of Lax operators
\begin{equation}
{\cal D_+}=\nabla_+\;-\;B_0,\;\;\;\;\;\;{\cal
D}_\DM=\nabla_\DM\;-\;A_0,\end{equation}
where
\begin{eqnarray}
B_0&=&{u\nabla_+\bar u\;-\;\bar u\nabla_+u\over 4u\bar u(1-u\bar u)}E,\\
A_0&=&{-1\over 4u\bar u(1-u\bar u)}\left(\begin{array}{cc}
(2u\bar u-1)(u\nabla_\DM\bar u-\bar u\nabla_\DM u)&4i\sqrt{u\bar u(1-u\bar
u)}\bar u\nabla_\DM u\nonumber\\ & & \\
{}~&~\\
4i\sqrt{u\bar u(1-u\bar u)}u\nabla_\DM\bar u & (1-2u\bar u)(u\nabla_\DM\bar
u-\bar u\nabla_\DM u)\\
\end{array}\right),\nonumber\end{eqnarray}
with
\begin{equation}
E=\left(\begin{array}{cc}
1 & 0\\
0&-1\\
\end{array}\right).
\end{equation}

Note that $B_0B_0=0$.

The equations of motion (12) emerge as integrability conditions of the
following equation
\begin{equation}
[\nabla_\DM\;-\;A_0,\,\nabla_+\;-\;B_0]=[\nabla_\DM,\nabla_+]=
2i\Sigma^+{\cal M}.\end{equation}
As a matter of fact, the (1,0) supersymmetric $SU(2)/U(1)$ coset model is
integrable and, hence, should possess an infinite number of (1,0)
super-covariant conserved currents. These currents can be constructed in a
number of ways. One can use the abelianization method of gauge connection
utilized in [5]. We will write explicitly only the local $U(1)$ current with
components
\begin{eqnarray}
J&=&{u\nabla_+\bar u\;-\;\bar u\nabla_+ u\over 1\;-\;u\bar u},\nonumber\\ & &
\\
\bar J&=&{u\nabla_\DM\bar u\;-\;\bar u\nabla_\DM u\over 1 \;-\;u\bar
u}.\nonumber\end{eqnarray}
It is easy to check that the $U(1)$ current satisfies the superspace
conservation law
\begin{equation}
\nabla_+\bar J\;+\;\nabla_\DM J=0.\end{equation}
All other nontrivial currents are nonlocal functions of $u,\;\bar u$.

Note that as implied by eq. (11), the theory can be thought of as the (1,0)
superspace generalization of the free Nambu-Goto string in four dimensional
Minkowski space [1].

\section{Supervortices}

{}~~~~In order to bring the model given by eq. (11) into correspondence with
the
theory of (super)vortices, we have to add to the action (11) an appropriate
potential of the scalar superfields $u,\;\bar u$. The point to be made is that
in
(1,0) superspace any superspace Lagrangian density has to carry a spinor
Lorentz
charge [9]. Since the superfields $u,\;\bar u$ are Lorentz scalars, one cannot
build up a proper (1,0) superspace potential in terms of $u,\;\bar u$ alone.
So, we are forced to introduce a new superfield carrying spinor Lorentz
charge. As such a superfield one can take a singlet $``-"$ spinor
supermultiplet
$\Psi_-$ [9]. This permits us to add the following term
\begin{equation}
S_{pert}={1\over2}\int d^3z
E^{-1}(\Psi_-\nabla_+\Psi_-\;+\;i\lambda\Psi_-\sqrt{1-u\bar u}),\end{equation}
where $\lambda$ is a constant. In spite of its somewhat bizarre form this (1,0)
superspace functional leads to the desired potential in component fields.
Indeed, in the sector of the auxiliary filed, the (1,0) superspace action (19)
gives rise to
\begin{equation}
\int d^2\sigma e^{-1}\;(-{1\over2}F^2\;-\;{1\over2}\lambda F\sqrt{1-u\bar u}),
\end{equation}
where
\begin{equation}
u\equiv u|,\;\;\;\bar u\equiv \bar
u|,\;\;\;iF\equiv\nabla_+\Psi_-|.\end{equation}
Elimination of the auxiliary field $F$ leads to the scalar potential
\begin{equation}
\int d^2\sigma e^{-1} \;{\lambda^2\over 8}(1\;-\;u\bar u).\end{equation}
Exactly this potential appears in the model of vortices in constant external
field [1, 5]. Note that the action in eq. (19) is not super-Weyl invariant due
to the latter term in eq. (19).

All in all, we can regard the following (1,0) locally supersymmetric action
\begin{equation}
S={ik\over4\pi}\int d^3E^{-1}{(\nabla_+u\nabla_\DM\bar u\;+\;\nabla_+\bar
u\nabla_\DM u)\over 1\;-\;u\bar u}\;+\;{1\over2}\int
d^3z E^-(\Psi_-\nabla_+\Psi_-\;+\;i\lambda\Psi_-\sqrt{1-u\bar u})\end{equation}
as a classical theory of supersymmetric vortices. The classical dynamics is
described by the following (1,0) super-covariant equations
\begin{eqnarray}
\nabla_+\Psi_-\;+\;{i\lambda\over 2}\sqrt{1-u\bar u}&=&0,\nonumber\\
\nabla_\DM \nabla_+ \bar u\;+\;{u\nabla_\DM\bar u\nabla_+\bar u\over 1-u\bar
u}\;+\;{\lambda\pi\over k}\Psi_-\bar u\sqrt{1-u\bar u}&=&0,\\
\nabla_\DM \nabla_+ u\;+\;{\bar u\nabla_\DM u\nabla_+ u\over 1-u\bar
u}\;+\;{\lambda\pi\over k}\Psi_-u\sqrt{1-u\bar u}&=&0,\nonumber\end{eqnarray}

It is quite amusing that in the (1,0) superspace formulation the potential in
(19) admits a natural geometrical interpretation. At the classical level, we
are allowed to make the following change of variables
\begin{equation}
\Psi_-\to\hat\Psi_-=\Psi_-\sqrt{1-u\bar u}.\end{equation}
Then, the action (19) takes the form
\begin{equation}
S_{pert}={1\over2}\int d^3z E^{-1}
\left({\hat\Psi_-\nabla_+\hat\Psi_-\over 1-u\bar
u}\;+\;i\lambda\;\hat\Psi_-\right).\end{equation}
In the given expression, the last term can be understood as a cosmological term
[9] of the (1,0) supergravity. The geometrical meaning of the first term in eq.
(26) becomes transparent if we add new terms to
eq. (26). Namely, instead of eq. (26) we can consider the following (1,0)
supersymmetrical expression
\begin{equation}
S_{pert}={1\over2}\int d^3z E^{-1}
\left({\hat\Psi_-\nabla_+\hat\Psi_-\over 1-u\bar
u}\;-\;{\hat\Theta_-\nabla_+\hat\Theta_-\over1-u\bar u}\;+\;{(\bar
u\nabla_+u\,-\,u\nabla_+\bar u)\over (1-u\bar u)^2}\hat\Psi_-\hat\Theta_-\;+\;
i\lambda\;\hat\Psi_-\right),\end{equation}
where we have introduced a new spinor superfield $\hat\Theta_-$.
Now the following action
\begin{eqnarray}
S(\lambda\to0)&=&{ik\over4\pi}\int d^3E^{-1}
{(\nabla_+u\nabla_\DM\bar u\;+\;\nabla_+\bar
u\nabla_\DM u)\over 1\;-\;u\bar u}\nonumber\\ & & \\
&+&{1\over2}\int d^3z E^{-1}\left({\hat\Psi_-\nabla_+\hat\Psi_-\over 1-u\bar
u}\;-\;{\hat\Theta_-\nabla_+\hat\Theta_-\over1-u\bar u}\;+\;{(\bar
u\nabla_+u\,-\,u\nabla_+\bar u)\over (1-u\bar u)^2}\hat\Psi_-\hat\Theta_-
\right).
\nonumber\end{eqnarray}
can be viewed as a (1,0) superspace realization of the
$SU(2)/U(1)$ coset with $N=2$ global supersymmetry. The additional
supertransformations written in terms of (1,0) superfields are as follows
\begin{eqnarray}
\delta_{(2,0)}u&=&\epsilon^+\nabla_+u,\nonumber\\
\delta_{(2,0)}\bar u&=&-\epsilon^+\nabla_+\bar u,\nonumber\\
\delta_{(2,0)}\hat\Psi_-&=&\epsilon^+\nabla_+\hat\Theta_-,\nonumber\\
\delta_{(2,0)}\hat\Theta_-&=&-\epsilon^+\nabla_+\hat\Psi_-,\nonumber\\
\delta_{(0,1)}u&=&\alpha\epsilon^-_1(\hat\Psi_-+\hat\Theta_-),\;\;\;\;\alpha=
\sqrt{2\pi/k}\nonumber\\
\delta_{(0,1)}\bar u&=&\alpha\epsilon^-_1(\hat\Psi_--\hat\Theta_-),
\nonumber\\ & & \\
\delta_{(0,1)}\hat\Psi_-&=&i\alpha^{-1}
\epsilon^-_1\nabla_\DM(u+\bar u),\nonumber\\
\delta_{(0,1)}\hat\Theta_-&=&i\alpha^{-1}
\epsilon^-_1\nabla_\DM(u-\bar u),\nonumber\\
\delta_{(0,2)}u&=&\alpha\epsilon^-_2(\hat\Psi_--\hat\Theta_-),\nonumber\\
\delta_{(0,2)}\bar u&=&-\alpha\epsilon^-_2(\hat\Psi_-+\hat\Theta_-),
\nonumber\\
\delta_{(0,2)}\hat\Psi_-&=&i\alpha^{-1}
\epsilon^-_2\nabla_\DM(u-\bar u),\nonumber\\
\delta_{(0,2)}\hat\Theta_-&=&-i\alpha^{-1}
\epsilon^-_2\nabla_\DM(u+\bar u),\nonumber
\end{eqnarray}
where $\epsilon^+$ is the Grassman parameter of the additional (1,0)
supersymmetry; $\epsilon^-_1,\;\epsilon^-_2$ are the Grassman parameters of the
two additional (0,1) supersymmetries, so that the transformations in (29) form
the (1,2) superalgebra\footnote{It is worthy to be emphasized that the $N=2$
(global) supersymmetry of the theory given by eq. (28) can be understood as a
general property of the two-dimensional supersymmetric non-linear sigma models
describing a complex target manifold [15, 16].}.

Thus, the model described by the action in eq. (23) can be realized as a (2,0)
reduction of the $N=2$ theory. We will exhibit the importance
of this fact in the course of the quantization of the theory under
consideration.

The $N=2$ supersymmetric action can be formulated as follows
\begin{eqnarray}
S&=&{ik\over4\pi}\int d^3E^{-1}
{(\nabla_+u\nabla_\DM\bar u\;+\;\nabla_+\bar
u\nabla_\DM u)\over 1\;-\;u\bar u}
\;+\;{1\over2}\int d^3z E^{-1}
(\Psi_-\nabla_+\Psi_-\;-\;\Theta_-\nabla_+\Theta_-\nonumber\\ & & \\
&+&{(\bar
u\nabla_+u\,-\,u\nabla_+\bar u)\over 1-u\bar u}\Psi_-\Theta_-
\;+\;i\lambda\Psi_-\sqrt{1-u\bar u}),\nonumber\end{eqnarray}
where
\begin{equation}
\hat\Theta=\Theta\sqrt{1-u\bar u}.
\end{equation}
This theory yields the following classical equations of motion
\begin{eqnarray}
\nabla_+\Psi_-\;-\;{1\over2}{(\bar u\nabla_+u-u\nabla_+\bar u)\over 1-u\bar
u}\Theta_-\;-\;{i\lambda\over 2}\sqrt{1-u\bar u}&=&0,\nonumber\\
\nabla_+\Theta_-\;+\;{1\over2}{(\bar u\nabla_+u-u\nabla_+\bar u)\over 1-u\bar
u}\Psi_-&=&0\nonumber\\ & & \\
\nabla_+\nabla_\DM\bar u\;+\;{(u\nabla_\DM\bar u+{i\pi\over k}\Psi_-\Theta_-)
\nabla_+\bar u\over 1-u\bar
u}\;+\;{\lambda\pi\over k}\Psi_-\bar u\sqrt{1-u\bar u}&=&0,\nonumber\\
\nabla_+\nabla_\DM u\;+\;{(\bar u\nabla_\DM u-{i\pi\over
k}\Psi_-\Theta_-) \nabla_+u\over 1-u\bar
u}\;+\;{\lambda\pi\over k}\Psi_-u\sqrt{1-u\bar u}&=&0,\nonumber\end{eqnarray}
As a matter of fact, this theory has a natural geometrical interpretation as an
$N=2$ supersymmetric coset perturbed by a cosmological constant-like term.

\section{Integrability}
{}~~~~It
is not inconceivable that the supersymmetric theory in eq. (30) continues to
be integrable. In order to clarify this property it is more convenient to
make use of the $N=2$ superfield formalism. Indeed, the theory in eq. (30)
admits a natural $N=2$ superspace formulation.

The component content of the theory (30) allows us to define two $N=2$ scalar
chiral superfields $\Phi$ and $\bar\Phi$ (about $D=2,\;N=2$ superspace see for
example [17]). Then the $N=2$ superspace action can be written as follows
\begin{equation}
S={k\over4\pi}\int d^6z E^{-1}\,K(\Phi,\bar\Phi)\;+\;\left(g\int
d^4z\mu\,\Phi+c.c.\right),\end{equation}
where in the kinetic term the K\"ahler potential $K$ is defined from the
equation
\begin{equation}
\partial_t K(t)={1\over t}\ln{1\over1-t},\end{equation}
with $t=\Phi\bar\Phi$. $\mu$ in the potential term of eq. (33) is a measure in
the curved chiral superspace which is a subspace of the curved $N=2$ superspace
with density $E^{-1}$. Note that the global restriction $u\bar u\le 1$ from
eqs. (10) amounts to the condition $t\le 1$ in eq. (34). The coupling $g$ is
given by
\begin{equation}
g={\sqrt{k\over 32\pi}}\lambda.\end{equation}

The $N=2$ theory yields the $N=2$ supercovariant equation of motion
\begin{equation}
\nabla^+\bar\nabla^+\partial_\Phi K(\Phi,\bar\Phi)=-g,\end{equation}
where $\nabla^+,\;\bar\nabla^+$ are $N=2$ supercovariant Grassman derivatives.
We can introduce a scalar superfield
$\bar S =-{1\over3}\partial_\Phi
K(\Phi,\bar\Phi)$. Then equation of motion (36) takes the form
\begin{equation}
\nabla^+\bar\nabla^+\bar S=g/3.\end{equation}

The very important point to be made is that the last equation can be observed
as an $N=2$ Liouville equation (without cosmological term)
\begin{equation}
\nabla^+\bar\nabla^+\bar S=-Q\hat R/3,\end{equation}
where $\bar S$ is a prepotential (superconformal factor) of the $N=2$
supergravity (see, e.g. [18]); whereas $\hat R$ is a constant $N=2$ scalar
super-curvature. This means that
the
$N=2$ geometry is no longer arbitrary in the problem under consideration. The
parameter $Q$ is to be fixed by the superconformal invariance [19].

Thus, we come to the following conclusion that the supersymmetrical extension
of the vortex
model of Lund and Regge leads naturally to the $N=2$ Liouville theory.
Therefore, the integrability of supervortices comes into being because of the
integrability of the Liouville theory.

\section{Quantization}

{}~~~~Now we would like to turn to some quantum aspects of the model under
consideration. We saw above that at the classical level we can get the
potential term (22) in two classically equivalent ways: either with superfield
action (30) or with superfield action (35). At the quantum level, these two
options are no longer indistinguishable. Indeed, the change of variables
proposed in eqs. (25), (31) entails the Jacobian which is nothing but the
chiral
anomaly of the corresponding (complex) Weyl spinor (built up of two (real)
Majorana-Weyl spinors $\Psi_-,\;\Theta_-$). Therefore, this Jacobian
will contribute to the functional integral over the scalar superfields
$u,\;\bar u$. At the same time we were convinced that the expression given by
eq. (35) provides more geometrical insight into the theory. So, we will
consider the (1,0) supersymmetric $SU(2)/U(1)$ coset perturbed by the term (27)
as a theory of supersymmetrical vortices. The classical action of the theory
is as follows
\begin{eqnarray}
S&=&{ik\over4\pi}\int d^3E^{-1}
{(\nabla_+u\nabla_\DM\bar u\;+\;\nabla_+\bar
u\nabla_\DM u)\over 1\;-\;u\bar u}
\;+\;{1\over2}\int d^3z E^{-1}
[
{\hat\Psi_-\nabla_+\hat\Psi_-\;-\;\hat\Theta_-\nabla_+
\hat\Theta_-\over 1-u\bar u}\nonumber\\ & & \\
&+&{(\bar
u\nabla_+u\,-\,u\nabla_+\bar u)\over (1-u\bar u)^2}\hat\Psi_-\hat\Theta_-
\;+\;i\lambda\hat\Psi_-].\nonumber\end{eqnarray}

To proceed with quantization of the given theory we note that in the conformal
limit $\lambda=0$ the action (28) describes the $N=2$ supersymmetrical coset.
As a matter of fact, this system has to be an exact finite conformal model. It
is actually the case, if one adds a suitable dilaton term to the classical
expression in eq. (28) [11]. Furthermore, the quasi-classical background
described by the classical supersymmetric action in (28) accompanied with a
proper dilaton is, in fact, exact in the quantum theory [12]. This property
is entirely due to supersymmetry. So, there are no quantum corrections to the
following supersymmetric sigma model
\begin{eqnarray}
S(\lambda&\to&0)={ik\over4\pi}\int d^3E^{-1}
{(\nabla_+u\nabla_\DM\bar u\;+\;\nabla_+\bar
u\nabla_\DM u)\over 1\;-\;u\bar u}\nonumber\\
&+&{1\over2}\int d^3z E^{-1}[ {\hat\Psi_-\nabla_+\hat\Psi_-\;-\;
\hat\Theta_-\nabla_+\hat\Theta_-\over 1-u\bar u} \;+\;{\bar
u\nabla_+u\,-\,u\nabla_+\bar u\over (1-u\bar u)^2}\hat\Psi_-\hat\Theta_-]\\
&-&{i\over8\pi}\int d^3z E^{-1}\ln (1-u\bar u)\Sigma^+,\nonumber
\end{eqnarray}
where the last term describes the dilaton. Thus, we come to conclusion that 2D
(super)gravity becomes very significant in (super)vortex theories.

When $\lambda\ne 0$, the last term in eq. (39) is just linear in the spinor
superfield. Therefore, it may give rise only to one-particle-reducible
diagrams. One can use the covariant (1,0) supergraphs [13, 14] to draw all
relevant diagrams in the theory. It is well known that one-particle-reducible
diagrams are finite if the composed one-particle-irreducible subdiagrams are
finite. Obviously all one-particle-reducible diagrams which may appear
in the theory given by eq. (39) have as their composed subdiagrams the
diagrams of the coset model. But all diagrams of the coset model are finite
provided a proper dilaton is taken into account. Hence, the theory in eq. (39)
with the dilaton term as in eq. (40) has to be finite beyond the conformal
limit.

Thus, as a theory of (1,0) supersymmetrical world sheet vortices we propose
the following model
\begin{eqnarray}
S&=&{ik\over4\pi}\int d^3z E^{-1}{(\nabla_+u\nabla_\DM\bar u\;+\;\nabla_+\bar
u\nabla_\DM u)\over 1\;-\;u\bar u}\nonumber\\
&+&{1\over2}\int d^3zE^{-1}\left(
{\hat\Psi_-\nabla_+\hat\Psi_-\over 1\;-\;u\bar u}\;-\;
{\hat\Theta_-\nabla_+\hat\Theta_-\over1-u\bar u}\;+\;{(\bar
u\nabla_+u\,-\,u\nabla_+\bar u)\over (1-u\bar u)^2}\hat\Psi_-\hat\Theta_-
\right)
\\
&+&{i\lambda\over 2}\int d^3z E^{-1}\hat\Psi_-
\;-\;{i\over8\pi}\int d^3zE^{-1}\ln(1-u\bar
u)\;\Sigma^+.\nonumber\end{eqnarray}

\section{Conclusion}

{}~~~~We believe that the results presented here justify the application
of the stringy ideas to (super)vortices on world sheet.  In the future
we plan to study in more depth the prospect of finiteness of the
models in the perturbative regime.  There is also the issues of
bosonization and whether further extend supersymmetry versions of this
theory exist.

\par \noindent
{\em Acknowledgement}: O. S. would like to thank E. Ramos and S. Thomas,
for discussions.  S.J.G. wishes to acknowledge the Mathematical Sciences
Research Institute of Berkeley, CA for hospitality extended during the
completion of this work.

\newpage


\begin{thebibliography}{66}

\bibitem{a} F. Lund and T. Regge, Phys. Rev. {\bf D14} (1976) 1524.
\bibitem{b} F. Lund, Phys. Rev. Lett. {\bf 38} (1977) 1175.
\bibitem{c} Y. I. Kogan, JETP Lett. {\bf 45} (1987) 709; B. Sathiapalan,
Phys. Rev. {\bf D35} (1987) 3277. \bibitem{d} B. A. Ovrut and S. Thomas, Phys.
Lett. {\bf B257} (1991) 292; Phys. Rev. {\bf D43} (1991) 1314.
\bibitem{e} I. Bakas, {\it Conservation laws and geometry of perturbed
coset models}, Preprint CERN-TH. 7047/93, hep-th/9310122.
\bibitem{f} H. J. De Vega and N. S\'anchez, {\it The general solution
of the 2-D sigma complex Sine-Gordon model}, Preprint PAR LPTHE 93/55,
hep-th/9312085.
\bibitem{g} K. Bardakci, E. Rabinovici and B. Saring, Nucl.
Phys. {\bf B299} (1988) 157; K. Gawedzki and A. Kupianen, Phys. Lett.
{\bf B215} (1988) 119; Nucl. Phys. {\bf B320} (1989) 625; D. Karabali,
Q.-H. Park, H. J. Schnitzer and Z. Yang, Phys. Lett. {\bf B216} (1989) 307.
\bibitem{h} S. J. Gates Jr., S. V. Ketov, S. M. Kuzenko and O. A. Soloviev,
Nucl. Phys. {\bf B362} (1991) 199-231.
\bibitem{i} R. Brooks, F. Muhammad and S. J. Gates Jr., Nucl. Phys. {\bf B268}
(1986) 599.
\bibitem{j} I. Bakas, {\it $W_\infty$ symmetry of the Nambu-Goto string in 4
dimensions}, Preprint CERN-TH. 7046, hep-th/9310121.
\bibitem{k} E. Witten, Phys. Rev. {\bf D44} (1991) 314;
R. Dijkgraaf, H. Verlinde and E. Verlinde, Nucl. Phys. {\bf B371} (1992) 269.
\bibitem{l} I. Jack, D. R. T. Jones and J. Panvel, Nucl. Phys. {\bf B393}
(1993) 95; A. A. Tseytlin,{\it Conformal sigma models corresponding to gauged
Wess-Zumino-Witten theories}, Preprint CERN-TH. 6804/93, hep-th/9302083.
\bibitem{m} S. J. Gates Jr., M. T. Grisaru, L. Mezincescu and P. K. Townsend,
Nucl. Phys. {\bf B286} (1987) 1.
\bibitem{n} S. V. Ketov and O. A. Soloviev, Int. J. Mod. Phys. {\bf A6} (1991)
2971; S. V. Ketov, S. M. Kuzenko and O. A. Soloviev, Class. Quantum Grav.
{\bf 7} (1990) 1403.
\bibitem{o} A. Sen, Nucl. Phys. {\bf B278} (1986) 289;
H. W. Braden, Nucl. Phys. {\bf B291} (1987) 516;
I. L. Buchbinder, S. M. Kuzenko and O. A. Soloviev, {\it The proper-time method
in two-dimensional superspace: The effective action and anomalies for
(non-linear) $(p,q)~\sigma$-models in (1,0) superspace}, Tomsk Preprint No. 53,
Tomsk (1988);
C. M. Hull and B. Spence, Nucl. Phys. {\bf B345} (1990) 493;
C. M. Hull, Mod. Phys. Lett. {\bf A5} (1990) 1793.
\bibitem{16} Y. Kazama and H. Suzuki, Nucl. Phys. {\bf B321} (1989) 232.
\bibitem{17} S. J. Gates Jr., M. T. Grisaru, M. Rocek and W. Siegel,
Superspace, or one thousand and one lessons in supersymmetry
(Benjamin-Cummings, Menlo Park, CA, 1983).
\bibitem{18} A. Alnowaiser, Class. Quantum Grav. {\bf 7} (1990) 1033.
\bibitem{18} A. M. Polyakov, Mod. Phys. Lett. {\bf A2} (1987) 893;
V. G. Knizhnik, A. M. Polyakov and A. B. Zamolodchikov, Mod. Phys. Lett. {\ A3}
(1988) 819.

\end{thebibliography}
\end{document}